\documentclass{aa}  
\usepackage{graphicx}

\begin{document}
\thesaurus{}
\title{Relativistic theory for time and frequency transfer\\ to order $c^{-3}$}
\author{Luc Blanchet\inst{1}, Christophe Salomon\inst{2}, Pierre Teyssandier\inst{3} 
and Peter Wolf\inst{4}}
\offprints{L. Blanchet}
\institute{D\'epartement d'Astrophysique Relativiste et de Cosmologie,
       Observatoire de Paris, 92195 Meudon Cedex, France
\and Laboratoire Kastler Brossel,
 Ecole Normale Sup\'erieure, 24 rue Lhomond 75231 Paris, France
\and Observatoire de Paris, DANOF, CNRS/UMR 8630,
61 avenue de l'Observatoire, 75014 Paris, France
\and Bureau International des Poids et Mesures,
Pavillon de Breteuil, F-92312 S\`evres Cedex, France}
\date{}
\titlerunning{relativistic time and frequency transfer}
\authorrunning{L. Blanchet, C. Salomon, P. Teyssandier and P. Wolf}
\maketitle

\begin{abstract}
This paper is motivated by the current development of several space
missions (e.g. ACES on International Space Station) that will use
Earth-orbit laser cooled atomic clocks, providing a time-keeping
accuracy of the order of $5\times 10^{-17}$ in fractional
frequency. We show that to such accuracy, the theory of frequency
transfer between Earth and Space must be extended from the currently
known relativistic order $1/c^2$ (which has been needed in previous
space experiments such as GP-A) to the next relativistic correction of
order $1/c^3$. We find that the frequency transfer includes the first
and second-order Doppler contributions, the Einstein gravitational
red-shift and, at the order $1/c^3$, a mixture of these effects. As
for the time transfer, it contains the standard Shapiro time delay,
and we present an expression also including the first and second-order
Sagnac corrections. Higher-order relativistic corrections, at least
${\cal O}(1/c^4)$, are numerically negligible for time and frequency
transfers in these experiments, being for instance of order $10^{-20}$
in fractional frequency. Particular attention is paid to the problem
of the frequency transfer in the two-way experimental
configuration. In this case we find a simple theoretical expression
which extends the previous formula (Vessot et al. 1980) to the next
order $1/c^3$. In the Appendix we present the detailed proofs of all
the formulas which will be needed in such experiments.
\end{abstract}

\section{Introduction}
 
Recent advances in laser cooling of atoms have led to the development
of a number of highly accurate atomic clocks (Caesium and Rubidium
fountains) which have improved time-keeping accuracy by an order of
magnitude during the nineties to currently $\approx 10^{-15}$ in
fractional frequency (Laurent et al. 1999, Bize et al. 1999). For
further improvement of accuracy, the fountain clocks are limited by
gravity, and therefore several experiments are planned for the near
future that will fly laser cooled atomic clocks on board terrestrial
satellites. One of these is ESA's Atomic Clock Ensemble in Space
(ACES) mission (Salomon \& Veillet 1996) planned for 2005. This
mission will place a laser cooled caesium clock together with a
hydrogen maser on board the International Space Station. When combined
with equally accurate time/frequency transfer systems that allow the
comparison between space and ground clocks, the experiment should
become a useful tool for a number of applications in metrology,
fundamental physics, atmospheric studies, geodesy etc.

Therefore the ACES mission will include sufficiently stable optical
and microwave time and frequency transfer systems to allow the
comparison of the clocks with negligible noise contribution from the
transfer system itself. At the required accuracy this condition calls
for two-way systems that exchange electromagnetic pulses in two
directions in order to eliminate or reduce a number of unwanted
effects associated with the instrumental delays, the propagation
within the ionosphere and troposphere, etc. The Time Transfer by Laser
Light (T2L2) system (Fridelance, Samain \& Veillet 1996) uses optical
pulses that are emitted on the ground, reflected by the satellite and
received back on the ground. The events of emission and reception are
dated on the ground clock, and the event of reflection on the
satellite clock.  The microwave system is expected to exchange pulses
in both directions and date all events of emission and reception. It
will also measure the frequency of the emitted and received pulses on
board and on the ground. Additionally it is expected to include two
down-link signals at different frequencies in order to allow the
cancellation of the residual ionospheric effect.

At the required uncertainties a fundamentally relativistic modelling
of the experiment is indispensable. Indeed the relativistic effects
when comparing two clocks separated by an altitude of $\approx 400$ km
(as is the case for ACES) can amount to several parts in $10^{-11}$,
which exceeds the expected uncertainties of the clocks ($\approx
5\times 10^{-17}$) by several orders of magnitude. Evidently, a
relativistic treatment of distant clock comparisons using
electromagnetic signals becomes necessary whenever the uncertainties
of such experiments become smaller than the size of the relativistic
corrections. This has already been the case for the gravity probe A
(GP-A) experiment (Vessot et al. 1980), and is more generally the case
since the advent of the Global Positioning System (GPS) which is now
extensively used for distant clock comparisons.

The first theoretical treatments of such comparisons in a relativistic
framework were performed by Jaffe \& Vessot (1976), Ashby \& Allan
(1979), Vessot, Levine et al. (1980), Allan \& Ashby (1985), and
Klioner (1993). Subsequent new methods with improved uncertainties,
such as the Two Way Time Transfer (TWTT) (Hetzel \& Soring 1993) and
the Laser Synchronization from Stationary Orbit (LASSO) (Veillet \&
Fridelance 1993), have led to the need for more accurate theoretical
treatments that include some higher order terms (Petit \& Wolf 1994,
1997, Wolf
\& Petit 1995, Klioner \& Fukushima 1994). The latter papers treat in
some detail the synchronization (time transfer) between distant clocks
using electromagnetic signals and the relation between the proper time
of a clock on the Earth or on board terrestrial satellites and the
geocentric coordinate time, TCG. The relativistic theory for frequency
transfer has been revisited recently by Ashby (1998), leading to
results similar to those presented in Subsection \ref{4.2}. In the
perspective of the expected uncertainties of the ACES and other
similar experiments, for time as well as frequency transfer, a
re-examination of these formalisms has proved necessary. This is the
subject of the present paper.

Essentially we shall compute the relativistic transfers of time and
frequency, including all the terms up to the order $1/c^3$. The
coordinate time transfer up to this order is well known as it consists
of the standard Shapiro time delay. Concerning the frequency transfer,
we find that the formula for the one-way transfer is rather
complicated. Our main result will concern the formula for the two-way
frequency transfer, up to order $1/c^3$. This formula appeared
previously in the paper by Ashby (1998) but without a detailed
derivation. In this paper, we derive this formula and, more generally,
we present a self-contained derivation of all the formulae needed in
this context. These formulae will be of direct use in the ACES
experiment, and {\it a fortiori} in more precise future
experiments. Higher-order relativistic corrections are negligible with
respect to the projected uncertainties associated with ACES.

The numerical applications made in this paper concern the ACES mission
with a transfer from the Space Station $A$ orbiting at the altitude
$H=400$ km to a ground station located at $B$. For the velocities
involved we use $v_A=7.7\, 10^3$ m/s and $v_B=v_{\rm ground}=465$ m/s;
for the gravitational potentials, $U_B/c^2= 6.9\, 10^{-10}$ and
$U_A/c^2= 6.5\, 10^{-10}$; and for the Earth parameters $GM_{\rm
E}=3.98\, 10^{14}$ ${\rm m}^3/{\rm s}^2$ and $R_{\rm E}=6.37\, 10^6$
m. We consider that the experimental uncertainties of ACES will be at
the level of 5 ps for time transfer and $5\times 10^{-17}$ for
frequency transfer.

The paper is organized in such a way that the main text summarizes all
the results needed by an experimental team in setting up a
relativistic time and/or frequency transfer, with all the proofs and
theoretical details relegated to the Appendix. In Section \ref{sect2}
we treat the problem of the transformation from proper time to
coordinate time TCG at a level sufficient for ground clocks and the
ACES space clock. In Section \ref{sect3} we give the expressions
required for time transfer (including Sagnac terms), and in Section
\ref{sect4} for frequency transfer. Both the one-way and two-way
transfers are considered in each case.

\section{Proper time in terms of coordinate time}\label{sect2}

Throughout this work we use the geocentric inertial (non-rotating)
coordinate frame GRS: Geocentric Reference System. Thus, $x^0/c=t=$TCG
is the geocentric coordinate time\footnote{The relation between
$t$ and the terrestrial time ${\rm TT}$ (realized by the International
Atomic Time, TAI) is given by $dt/d{\rm TT}=1+L_g$ where $L_g$ is a
defining constant fixed in IAU resolution B1.9 (IAU 2000) as
$L_g=6.969290134~10^{-10}$; this value was chosen to be close to the
previous definition $L_g=W_0/c^2$ (IAU 1991) where $W_0$ is the Earth
potential (gravitational plus centrifugal) at the reference geoid
close to the mean surface of the oceans.}, and ${\bf x}=(x^i)$ are the
GRS harmonic spatial coordinates, for which the spatial metric
is conformally flat to order $1/c^2$. In these coordinates the
metric interval including all the terms up to the order $1/c^2$ reads
 
\begin{eqnarray}\label{1}
ds^2&\equiv& -c^2 d\tau^2\nonumber\\ &=&-\left(1-{2U\over
c^2}\right)c^2 dt^2+\left(1+{2U\over c^2}\right)\delta_{ij} dx^idx^j\;,
\end{eqnarray}
where $U$ is the total Newtonian potential, with the convention that
$U\geq 0$ (IAU 1991), and where $\delta_{ij} dx^idx^j=d{\bf x}^2$
denotes the Euclidean space metric. The Solar-system barycentric
coordinates, centred on the barycenter of the Solar system, are
denoted ${\cal T}=$TCB: solar system barycentric coordinate time, and
${\cal X}^i$: solar system spatial (harmonic) coordinates. The
proper time of a clock $A$ located at the GRS coordinate position
${\bf x}_A(t)$, and moving with the coordinate velocity ${\bf
v}_A=d{\bf x}_A/dt$, is

\begin{eqnarray}\label{2}
{d\tau_A\over dt}=1&-&{1\over c^2}\biggl[{{\bf v}_A^2\over 2}+
U_{\rm E}({\bf x}_A)\nonumber\\ &&\quad+V({\cal X}_A)-V({\cal
X}_{\rm E})-x_A^i \partial_iV({\cal X}_{\rm E})\nonumber\\
&&\quad+x_A^i Q_i\biggr]\;.
\end{eqnarray}
Here, $U_{\rm E}$ denotes the Newtonian potential of the Earth at the
position ${\bf x}_A$ of the clock in the GRS frame, and $V$ is the sum
of the Newtonian potentials of the other bodies (mainly the Sun and
the Moon), either at the position ${\cal X}_{\rm E}$ of the Earth
center of mass in barycentric coordinates, or at the clock
location ${\cal X}_{\rm A}$. The three terms involving the potential
$V$ represent the tidal field of the other bodies at the position of
the clock. [To a good approximation (introducing errors of much less
than $10^{-18}$) one can write $V({\cal X}_{\rm A}) \approx V({\cal
X}_{\rm E}+{\bf x}_A)$. Then, as usual, the tidal field at the
position ${\bf x}_A$ can be approximated using a Taylor expansion by
the standard expression $\frac{1}{2}x_A^ix_A^j\partial_{ij}V({\cal
X}_{\rm E})$ again neglecting terms smaller than $10^{-18}$.] The
terms due to the tides of the other bodies (i.e. involving $V$) are
small for the ground station and the low orbit of ACES (of order
$2\times 10^{-17}$) and are either negligible or easily evaluated if
required (depending on the final clock accuracies reached). The last
term in (\ref{2}) is due to the non-geodesic acceleration of the
center of mass of the Earth that is induced by the mass quadrupole of
the Earth\footnote{To a sufficient approximation we have

\begin{eqnarray*}
Q_i&=&-{1\over 2M_{\rm E}} I_{\rm E}^{jk}\partial_{ijk}V({\cal X}_{\rm
E})\;,
\end{eqnarray*}
where $I_{\rm E}^{jk}$ is the Earth quadrupole moment (see
e.g. Brumberg \& Kopejkin 1990 for details).}, and which is
numerically of the order of $3\times 10^{-11} {\rm m/s^2}$. This term is
negligible in the case of the envisioned experiments: numerically
$|Q_i x_A^i/c^2|$ amounts to less that $10^{-20}$.

Thus, for application to ACES, and more generally to any
experiment at a level of uncertainty greater than $5\times 10^{-17}$ on a
satellite at similar altitude as ACES, we can keep only the first
three terms in the relation (\ref{2}) between the proper time $\tau_A$
and the coordinate time $t$:
 
\begin{equation}\label{3}
{d\tau_A\over dt}=1-{1\over c^2}\left[{{\bf v}_A^2\over 2}+U_{\rm
E}({\bf x}_A)\right]\;.
\end{equation}
Note that at this level of uncertainty it is crucial to take into
account in the relation (\ref{3}) the non-sphericity (oblaticity) of
the Earth Newtonian potential. In fact, it is even not sufficient to
model the Earth potential with a $J_2$-term taking into account the
quadrupolar deformation. Rather, the potential $U_{\rm E}({\bf x}_A)$
should be computed according to the detailed procedures of Wolf and
Petit (1995), and Petit and Wolf (1997). For example, for a clock $B$
fixed on the Earth surface, the relativistic correction term appearing
in the coordinate/proper time relation (\ref{3}) is given with the
required precision by

\begin{equation}\label{4}
{{\bf v}_B^2\over 2}+U_{\rm E}({\bf x}_B)=W_0-\int_0^{H_B} \!g~dH\;,
\end{equation}
where $W_0$ is the Earth potential at the reference geoid
($W_0=62636856 ~{\rm m^2/s^2}$), where $g$ denotes the (gravitational
plus centrifugal) Earth acceleration, and where $H_B$ is the geometric
height of the clock above the reference geoid. Note that this
procedure is limited by the uncertainty in the determination of $W_0$
which at present gives rise to an error of about 1 part in $10^{17}$
in $d\tau_B/dt$. For uncertainties below that level, it is expected
that, inversely, clock comparisons with highly accurate space clocks
will yield the best estimate of the potential on the ground.
 
\section{Coordinate time transfer}\label{sect3}

\subsection{One-way signal transmission}

Let $A$ be the emitting station, with GRS position ${\bf x}_A(t)$, and
$B$ the receiving station, with position ${\bf x}_B(t)$. We denote by
$t_A$ the coordinate time at the instant of emission of a light
signal, and by $t_B$ the coordinate time at the instant of
reception. We put $r_A=|{\bf x}_A(t_A)|$, $r_B=|{\bf x}_B(t_B)|$ and
${\rm R}_{AB}=|{\bf x}_B(t_B)-{\bf x}_A(t_A)|$, where $|~|$ is the
Euclidean norm associated with the metric $\delta_{ij}$.  Up to the
order $1/c^3$ the coordinate time transfer $T_{AB}\equiv t_B-t_A$ is
given by

\begin{equation}\label{6}
T_{AB}={{\rm R}_{AB}\over c}+{2GM_{\rm E}\over
c^3}\ln\left({r_A+r_B+{\rm R}_{AB}\over r_A+r_B-{\rm
R}_{AB}}\right)\;,
\end{equation}
where the logarithmic term represents the Shapiro time
delay\footnote{In the case of a general metric theory of gravity, the
factor two in front of the Shapiro time delay should be replaced by
$1+\gamma$ where $\gamma$ is the standard PPN parameter.} (Shapiro
1964). See the Appendix for several derivations of the Shapiro time
delay and for an alternative expression given by (\ref{a32}).  In the
case of zenithal geometry, i.e. propagation of the signal along the
local vertical (for which $|{\bf x}_B-{\bf x}_A|=|r_B-r_A|$), between
the orbit of ACES at 400 km (assumed in all numerical examples below)
and the ground, the Shapiro time delay is 2 ${\rm ps}$. In
the case of zero elevation, it is 11 ${\rm ps}$.

In a real experiment, the position of the receptor $B$ may be recorded
at the time of emission $t_A$ rather than at the time of reception
$t_B$, i.e. we may have more direct access to ${\bf x}_B(t_A)$ rather
than ${\bf x}_B(t_B)$, and the formula (\ref{6}) gets modified by some
Sagnac correction terms consistently to the order $1/c^3$. In this
case the formula becomes

\begin{eqnarray}\label{7}
T_{AB}&=&{{\rm D}_{AB}\over c}+{{\bf D}_{AB}.{\bf v}_B(t_A)\over
c^2}\nonumber\\ &+&{{\rm D}_{AB}\over 2c^3}\left({\bf v}_{B}^2+{({\bf
D}_{AB}.{\bf v}_B)^2\over {\rm D}_{AB}^2}+{\bf D}_{AB}.{\bf
a}_B\right)\nonumber\\ &+&{2GM_{\rm E}\over c^3}\ln\left({r_A+r_B+{\rm
D}_{AB}\over r_A+r_B-{\rm D}_{AB}}\right)\;,
\end{eqnarray}
where ${\bf D}_{AB}={\bf x}_B(t_A)-{\bf x}_A(t_A)$ is the
``instantaneous'' coordinate distance between $A$ and $B$ at the
instant of emission at $A$ (we have ${\rm D}_{AB}=|{\bf D}_{AB}|$),
where ${\bf v}_B(t_A)$ denotes the coordinate velocity of the station
$B$ at that instant, and where ${\bf a}_B$ is the acceleration of $B$
(in all the small correction terms of order $1/c^3$ one can use,
independent of the required order, the quantities at the instants
$t_A$ or $t_B$). The second term in (\ref{7}) represents the Sagnac
term of order $1/c^2$ and can amount to 200 ${\rm ns}$ at low
elevation; the third term or Sagnac of order $1/c^3$ is about 5 ${\rm
ps}$ at low elevation (to be compared with the Shapiro term, which is
11 ${\rm ps}$).

\subsection{Two-way time transfer}

\begin{figure}
\includegraphics[width=8cm]{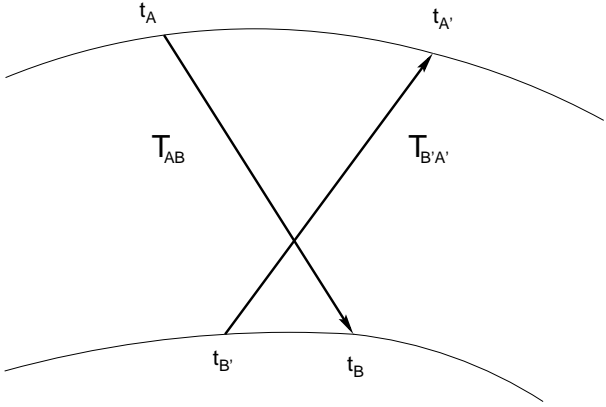}
\caption{Two-way time transfer in the non-rotating frame}\label{fig1}
\end{figure}
One signal is emitted from the satellite $A$ at instant $t_A$ and
received by the ground station $B$ at instant $t_B$. A second signal
(prime) is emitted by the ground $B$ at instant $t_{B'}$ and received
on $A$ at $t_{A'}$ (see Figure \ref{fig1}). The two transmission times
are

\begin{eqnarray}\label{8}
T_{AB}&=&t_B-t_A\;,\\
T_{B'A'}&=&t_{A'}-t_{B'}\;.
\end{eqnarray}
Furthermore, we denote the intervals of time between emission and
reception on board the satellite $A$ and at the station $B$ by

\begin{eqnarray}\label{9}
t_{AA'}&=&t_{A'}-t_A\;,\\
t_{B'B}&=&t_B-t_{B'}\;.
\end{eqnarray}
From these definitions we deduce the quantity $\Delta t$ which is
required for synchronization, namely

\begin{equation}\label{10}
\Delta t=t_A-t_{B'}={1\over 2}\left(t_{B'B}-t_{AA'}
+T_{B'A'}-T_{AB}\right) \;.
\end{equation}
Notice that $\Delta t$ is known because it is expressed in terms of
the two transmission times $T_{AB}$ and $T_{B'A'}$ which are {\it
computed} from the theoretical formulas (\ref{6}) or (\ref{7}) valid
for the one-way time transfer, and in terms of the time intervals
$t_{AA'}$ and $t_{B'B}$ which are {\it measured} on the satellite and
on the ground respectively\footnote{In the case of the time transfer
by laser light T2L2 to be operated on ACES we have $t_{AA'}=0$ since
the signal is reflected instantaneously.}.

\section{Frequency Transfer}\label{sect4}

\subsection{One-way transfer}
 
The frequency transfer between two clocks requires the determination of
the ratio $f_A/f_B$ between the proper frequencies $f_A$ and $f_B$
delivered by the clocks on the satellite ($A$) and on the ground
($B$). In practice this is achieved using a transmission of photons
from $A$ to $B$ and the formula

\begin{equation}\label{11}
{f_A \over f_B}=\left({f_A \over \nu_A}\right)\left({\nu_A \over
\nu_B}\right)\left({\nu_B \over f_B}\right)\;,
\end{equation}
where $\nu_A$ is the proper frequency of the photon as measured on $A$
(instant of emission $t_A$), and $\nu_B$ the proper frequency of the
same photon on $B$ at $t_B$. The first bracket in (\ref{11}) is
measured on $A$, the second bracket is given by the theoretical
formula (\ref{12}) below, and the third one is measured on $B$. In the
one-way transfer of photons, as derived in the Appendix, we have

\begin{equation}\label{12}
{\nu_A\over\nu_B}={1-{1\over c^2}\left[U_{\rm E}({\bf r}_B)+{{\bf
v}_B^2\over 2}\right]\over 1-{1\over c^2}\left[U_{\rm E}({\bf
r}_A)+{{\bf v}_A^2\over 2}\right]}~{q_A \over q_B}\;.
\end{equation}
For convenience in the notation, we henceforth denote the radial
vectors (in GRS coordinates) by ${\bf r}_A={\bf x}_A(t_A)$ and ${\bf
r}_B={\bf x}_B(t_B)$; and, as before, we have $r_A=|{\bf r}_A|$ and
$r_B=|{\bf r}_B|$, as well as the coordinate velocities ${\bf
v}_A={\bf v}_A(t_A)$ and ${\bf v}_B={\bf v}_B(t_B)$. To the required
order $1/c^3$, the last factor in (\ref{12}) is obtained from

\begin{eqnarray}
q_A=1&-&{{\bf N}_{AB}.{\bf v}_A\over c}\nonumber\\ &-&{4G M_{\rm
E}\over c^3}{(r_A+r_B){\bf N}_{AB}.{\bf v}_A+{\rm R}_{AB}{{\bf
r}_A.{\bf v}_A\over r_A}\over (r_A+r_B)^2-{\rm
R}_{AB}^2}\;,\label{13a}\\ q_B=1&-&{{\bf N}_{AB}.{\bf v}_B\over
c}\nonumber\\ &-&{4G M_{\rm E}\over c^3}{(r_A+r_B){\bf N}_{AB}.{\bf
v}_B-{\rm R}_{AB}{{\bf r}_B.{\bf v}_B\over r_B}\over (r_A+r_B)^2-{\rm
R}_{AB}^2}\;,\label{13b}
\end{eqnarray}
with ${\bf R}_{AB}={\bf r}_B-{\bf r}_A$, ${\rm R}_{AB}=|{\bf R}_{AB}|$
and ${\bf N}_{AB}={\bf R}_{AB}/{\rm R}_{AB}$. See the Appendix for the
derivation of these formulas. Note that the result
(\ref{13a})-(\ref{13b}) has been obtained in the Appendix assuming
that the field of the Earth is spherically symmetric. Indeed, the
$J_2$-terms in the factor $q_A/q_B$ do not exceed $4\times 10^{-17}$.

In the case of ACES, the various contributions in the one-way
frequency transfer (\ref{12})-(\ref{13b}) are numerically as follows.
First-order Doppler effect: for the satellite $|{\bf N}_{AB}.{\bf
v}_A/c| \leq 2.6\times 10^{-5} $; for the ground $|{\bf N}_{AB}.{\bf
v}_B/c| \leq 1.6\times 10^{-6}$. Second-order Doppler effect: ${\bf
v}_A^2/(2 c^2) \leq 3.4\times 10^{-10}$ for the satellite; ${\bf
v}_B^2/(2 c^2) \leq 1.3\times 10^{-12}$ for the ground. Gravitational
red-shift (Einstein) effect: $U_A/c^2\equiv U_{\rm E}({\bf
r}_A)/c^2=6.5 \times 10^{-10}$; $U_B/c^2=6.9 \times 10^{-10}$. The
terms of order $1/c^3$ are less than $3.6\times 10^{-14}$ for the
satellite and $2.2\times 10^{-15}$ for the ground.

\subsection{Two-Way frequency transfer}\label{4.2}

\begin{figure}
\includegraphics[width=8cm]{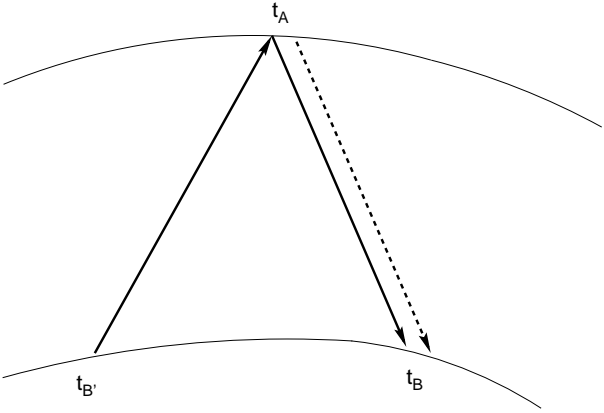}
\caption{Two-way frequency transfer in the non-rotating frame.}
\label{fig2}
\end{figure}
A ``tracking'' signal is sent from the ground station $B$ at instant
$t_{B'}$, received on the satellite $A$ at instant $t_A$ and
instantaneously re-emitted by a satellite transponder toward $B$ where
it is received at instant $t_B$. The down-link ``clock'' signal is
emitted simultaneously with the tracking signal at the transponding
instant $t_A$, and received at $t_B$ (see Figure \ref{fig2}) (Vessot
et al. 1980).  In the two-way frequency transfer the ratio
$\nu_A/\nu_B$ in (\ref{11}), needed for frequency comparison, is given
by

\begin{equation}\label{14}
{\nu_B\over \nu_A}={1\over 2}\left({\nu_B\over
\nu_{B'}}\right)+\Delta_{AB}+{1\over 2}\;.
\end{equation}
The ratio $(\nu_B/\nu_{B'})=(\nu_B/\nu_{B'})_{\rm station}$ is {\it
measured} at the ground station B, while $\Delta_{AB}$ is {\it
computed} by means of the theoretical formula (derived in the
Appendix)

\begin{eqnarray}\label{15}
\Delta_{AB}\!&=&\!{1\over c^2}\!\left[U_{AB}-{1\over 2}{\bf v}_{AB}^2
-{\bf R}_{AB}.{\bf a}_B\right]\!\left(1+{{\bf N}_{AB}.{\bf
v}_{AB}\over c}
\right)\nonumber\\
&+&{{\rm R}_{AB}\over c^3}\left(-{\bf v}_A.{\bf a}_B+{\bf R}_{AB}.{\bf
b}_B+2{\bf v}_B.{\bf a}_B-{\bf v}_B.\nabla U_B\right)\;.\nonumber\\
\end{eqnarray}
The difference of potentials between ground and satellite reads
$U_{AB}=U_B-U_A$ where $U_A \equiv U_{\rm E}[{\bf r}_A(t_A)]$ and $U_B
\equiv U_{\rm E}[{\bf r}_B(t_B)]$ (note that $U_{AB}>0$); the gradient
is $\nabla U_B=\partial U_{\rm E}({\bf r}_B)/\partial {\bf r}_B$; the
relative velocity is ${\bf v}_{AB}={\bf v}_A(t_A)-{\bf v}_B(t_B)$; the
acceleration of the ground is ${\bf a}_B={\bf a}_B(t_B)$; the
derivative of acceleration is ${\bf b}_B=d{\bf a}_B/dt$. 

In the case of ACES we find the following numerical contributions. For
the three dominant terms appearing at the order $1/c^2$,

\begin{eqnarray}\label{16}
&&\Delta_{\rm Einstein}={U_{AB}\over c^2}=4.6\, 10^{-11}\;;\\
&&\Delta_{\rm 2d-order~\! Doppler}=\Big|-{1\over 2c^2}{\bf
v}_{AB}^2\Big| \leq 3.3\, 10^{-10}\;;\\ &&\Delta_{\rm
acceleration}=\Big|-{1\over c^2}{\bf R}_{AB}.{\bf a}_B\Big|\leq 7\,
10^{-13}\;.
\end{eqnarray}
In the third term, or ``acceleration'' term, we have assumed that the
station $B$ is located at the equator, and we have used the upper
bound $({\rm R}_{AB})_{\rm max}~\omega_{\rm E}^2 ~\!r_B/c^2$, where
$\omega_{\rm E}$ is the Earth's angular velocity of rotation. Now,
according to (\ref{15}), in order to reach the $1/c^3$ precision, the
three previous terms are to be corrected by a factor that looks
exactly like a first-order Doppler effect. Numerically, we have

\begin{equation}\label{17}
\quad \bigg|{{\bf N}_{AB}.{\bf v}_{AB}\over c}\bigg|\leq 2.7 \, 10^{-5}\;,
\end{equation}
so this first-order Doppler factor induces a correction of the
frequency shift at the minimal level $8.2 \times 10^{-15}$, which is
measurable by ACES. Finally the four last terms in (\ref{15}) are
purely of order $1/c^3$. They amount respectively to the maximal
values $\leq 2\, 10^{-17}$, $3.5\, 10^{-19}$, and much less for the
last two terms. These last four terms are in general negligible for
ACES. The terms which are neglected in the formula (\ref{15}), which
are at least ${\cal O}(1/c^4)$, are numerically of the order of
$10^{-20}$ or less, too small to be detected by ACES.

In summary, the formula (\ref{15}) and its derivation in the Appendix
constitute the main results of this paper. To the dominant order
$1/c^2$ the expression was first derived by Vessot et al. (1980) who
used it in their GP-A experiment. To the order $1/c^3$ the formula
(\ref{15}) appears in a recent paper by Ashby (1998) but without
detailed derivation. The final expression (\ref{15}) is relatively
simple, as compared, for instance, to the corresponding expression
valid for the one-way frequency transfer [see
(\ref{12})-(\ref{13b})]. We find indeed that many of the scalar
products between ${\bf N}_{AB}$ and some velocities, though present
in the intermediate steps of the calculation, drop out from the
final result. The exception is for the factor in (\ref{15}) which is
made of the combination $1+{\bf N}_{AB}.{\bf v}_{AB}/c$ and which can
be nicely interpreted as a modification, at the level $1/c^3$, of the
dominant term of order $1/c^2$ by a first-order Doppler effect. The
set of equations for time and frequency transfers given in this paper
should be sufficient for the analysis of the planned clock experiments
in Earth orbit, at the level $5\times 10^{-17}$ in fractional frequency.

\appendix

\section{Theory}
 
This Appendix presents several equivalent derivations of the formulas
for the time and frequency transfers in the ACES experiment. Some of
the basic material needed in these derivations is not new and can be
found in standard textbooks such as Misner, Thorne \& Wheeler (1973),
and Will (1981). On the other hand, the problem of the propagation of
light in a gravitational field has been solved in a general way at the
linearized approximation: see Kopeikin \& Sch\"afer (1999) for a
complete investigation and an entry to the literature. Here we use the
explicit solution of the photon motion, the optical distance function
for stationary space-times (Buchdahl 1970, 1979), and the
differentiation of the well-known Shapiro (1964) formula.

We shall first consider the transfer of {\it coordinate} geocentric
time $t$; for this purpose it is sufficient to approximate the
gravitational field of the Earth as spherically symmetric (monopolar),
and to neglect tidal terms, hence $U_{\rm E}=G M/r$. However, as we
have seen in Section \ref{sect2}, higher spherical harmonics in the
Earth potential are needed in the relation between coordinate and
proper time. We denote the mass of the Earth by $M=M_{\rm E}$, the
Cartesian geocentric coordinates by $(t,{\bf r})$, with $t=$TCG and
${\bf r}={\bf x}$ agreeing to this approximation with the geocentric
spatial coordinates ($r=|{\bf r}|$). Using spherical coordinates
$\{t,r,\theta,\varphi\}$ associated in the standard way to the
geocentric coordinates ($0\leq\theta\leq\pi$, $0\leq\varphi\leq
2\pi$), we can write the 1PN metric as

\begin{equation}\label{a1}
ds^2=-f(r)c^2 dt^2+g(r)\left[dr^2+r^2d\theta^2+r^2 \sin^2\theta
d\varphi^2\right]\;,
\end{equation}
where to this order

\begin{eqnarray}
f(r)&=&1-{2U\over c^2}=1-{2G M\over rc^2}\;,\label{a2}\\
g(r)&=&1+{2U\over c^2}=1+{2G M\over rc^2}\;,\label{a3}
\end{eqnarray}
As said above, the monopolar approximation to the Earth potential is
sufficient for our purpose in most of this Appendix.

\noindent {\it Photon equations of motion.} 
In the geometric optics approximation, where the photon's wavelength
is much smaller than the typical size of the electromagnetic wave
packet, as well as of the space-time radius of curvature, the photon's
wave vector is null, $k_\mu k^\mu=0$, and is hypersurface-orthogonal,
$k_\mu=\partial_\mu S$. The differential equation for the trajectory
of the ray is $k^\mu=dx^\mu/dp=g^{\mu\nu}(x)\partial_\nu S(x)$. We
have $0=\nabla_\mu (k_\nu k^\nu)=2k^\nu\nabla_\mu k_\nu=
2k^\nu\nabla_\nu k_\mu$, where we have used the fact that $\nabla_\mu
k_\nu$ is symmetric as a consequence of $k_\mu=\partial_\mu S$, so we
find that $k^\nu\nabla_\nu k_\mu=0$, which is the geodesic equation, a
more useful form of which reads

\begin{equation}\label{a4}
{dk_\mu\over dp}={1\over 2}k^\rho k^\sigma\partial_\mu g_{\rho\sigma}\;.
\end{equation}
Here, $p$ denotes an affine parameter along the trajectory, and the
photon's null wave vector $k^\mu$ is such that

\begin{equation}\label{a5}
k^\mu={dx^\mu\over dp} \ ;\quad k_\mu=g_{\mu\nu} k^\nu \ ;\quad
0=k_\mu k^\mu\;,
\end{equation}
($k^\mu$ is future directed, $k^0=ck^t>0$). 

Although the metric (\ref{a1})-(\ref{a3}) is spherically-symmetric, it
is convenient to suppose that the motion of the photon takes place in
an arbitrary plane, not necessarily the equatorial plane
$\theta=\pi/2$.  For instance, we can consider that $\theta=0$
represents the Earth's rotation axis, and that the emitted signal
comes from a satellite moving on any orbit with inclination angle $i$
with respect to the equator. Since the metric is stationary and
axi-symmetric, we find immediately the first integrals $k_t=-{\hat E}$
and $k_\varphi={\hat L}$, where ${\hat E}$ and ${\hat L}$ denote two
constants along the trajectory (with ${\hat E}>0$ to ensure
$k^0>0$). Relating $k_t$ and $k_\varphi$ to the contravariant
components of the wave vector, $k^t=dt/dp$ and
$k^\varphi=d\varphi/dp$, yields the two integrals of motion

\begin{eqnarray}
\hat E&=&f(r)c^2{dt\over dp}\;,\label{a6a}\\
\hat L&=&g(r)r^2 \sin^2\theta{d\varphi\over dp}\;.\label{a6b}
\end{eqnarray}
Next, the equation corresponding to the $\theta$-coordinate reads
$dk_\theta/dp=g(r) r^2 \sin\theta\cos\theta(k^\varphi)^2$. Inserting
in this equation $k^\varphi=d\varphi/dp$ as deduced from (\ref{a6b}),
and noting that $g(r) r^2=g_{\theta\theta}$, we obtain
$d(k_\theta^2)/dp=2{\hat L}^2\cos\theta/\sin^3\theta d\theta/dp$. This
equation shows that $k_\theta$ is a function of the
$\theta$-coordinate only, and we get $k_\theta^2={\hat b}^2-{\hat
L}^2/\sin^2\theta$ where $\hat b$ is a new constant of the
motion. Hence,

\begin{equation}\label{a7}
r^4g^2(r)\left({d\theta\over dp}\right)^2={\hat b}^2-{{\hat
L}^2\over\sin^2\theta}\;.
\end{equation}
Finally, since $g^{\mu\nu}k_\mu k_\nu=0$ along a light ray,
we get from the previous integrals of motion,

\begin{equation}
g^2(r)\left({dr\over dp}\right)^2={g(r)\over f(r)}{{\hat E}^2\over c^2}
-{{\hat b}^2\over r^2}\;.
\end{equation}
It can be checked that the equation concerning the $r$-coordinate,
i.e.  $dk_r/dp=1/2k^\rho k^\sigma\partial_r g_{\rho\sigma}$, is now
automatically satisfied. Thus, the photon motion depends on the
constants $\hat E$, $\hat b$ and $\hat L$; however, by eliminating the
affine parameter $p$ in favor of the coordinate time $t$ we can
parametrize the motion by only two constants, say $b={\hat b}c/{\hat
E}$ and $L={\hat L}c/{\hat E}$. Finally, the solution reads

\begin{eqnarray}\label{a8}
{dr\over cdt}&=&\varepsilon_r {f\over g}\sqrt{{g\over f}-{b^2\over
r^2}}\;,\\ r^2{d\theta\over cdt}&=&\varepsilon_\theta {f\over
g}\sqrt{b^2-{L^2\over \sin^2\theta}}\;,\\ r^2{d\varphi\over
cdt}&=&{f\over g}{L\over \sin^2\theta}\;.
\end{eqnarray}
We have introduced the signs $\varepsilon_r$ and $\varepsilon_\theta$
of $dr/dt$ and $d\theta/dt$. Because the metric (\ref{a1}) is
spherically-symmetric, one could rotate the coordinates so that the
plane of the motion is simply the equatorial plane
$\theta=\pi/2$. This would correspond to $b_{\rm equatorial}=L$, and
in that case $\varphi$ would simply be the polar angle within the
orbital plane. In our more general situation, it is not difficult to
find the equation of the orbital plane. We introduce the angle
$\alpha$ defined by

\begin{equation}\label{a9}
\cos\alpha={b \cos\theta\over \sqrt{b^2-L^2}} \ ; 
\quad\sin\alpha=\varepsilon_\theta \sqrt{{b^2\sin^2\theta-L^2\over b^2-L^2}}\;.
\end{equation}
In terms of $\alpha$ we can integrate the equation for the azimuthal
angle $\varphi$ in the form

\begin{equation}\label{a10}
\tan (\varphi-\varphi_1)={b\over L}\tan\alpha\;,
\end{equation}
where $\varphi_1$ is an arbitrary constant. Clearly this is the
equation of the orbital plane. Its line of node is defined by the
direction $\varphi=\varphi_1-\pi/2$ in the equatorial plane, and the
inclination angle $i$ of the orbit is such that $\sin
i=\sqrt{1-L^2/b^2}$. The angle $\alpha$ is the polar angle in the
orbital plane, oriented in the sense of the motion (we have
$d\alpha^2=d\theta^2+\sin^2\theta d\varphi^2$). Introducing the
Euclidean orthonormal basis ${\bf e}_r$, ${\bf e}_\theta$, ${\bf
e}_\varphi$ associated with the spherical coordinates
$\{r,\theta,\varphi\}$, the position and velocity of the photon read

\begin{eqnarray}
{\bf r}&=&r {\bf e}_r\;,\label{a11a}\\ {d{\bf r}\over cdt}&\equiv&
{\bf n}\sqrt{f(r)\over g(r)}={dr\over cdt} {\bf e}_r+r {d\alpha\over
cdt} {\bf e}_\alpha\;,\label{a11b}
\end{eqnarray}
where ${\bf e}_\alpha$, namely the unit vector in the direction of
increasing $\alpha$ within the orbital plane, is given by

\begin{equation}\label{a12}
{\bf e}_\alpha=\varepsilon_\theta {\bf e}_\theta \sqrt{1-{L^2\over b^2
\sin^2\theta}}+{\bf e}_\varphi {L\over b \sin\theta}\;.
\end{equation}
In (\ref{a11b}) we have introduced the unit tangent vector ${\bf n}$
along the photon's path, which is directly related to the {\it
contravariant} components of the wave vector: $n^i=k^i/|{\bf k}|$,
where $|{\bf k}|$ is the Euclidean norm, so that ${\bf
n}^2=\delta_{ij}n^in^j=1$. From the facts that $k^\mu$ is a null
vector and that the line element (\ref{a1}) is diagonal and spatially
conformally flat, we can check that $|{\bf k}|=\sqrt{f/g}~k^0$, from
which we deduce that the {\it covariant} components $k_i$ are also
proportional to $n^i$, {\it viz}

\begin{equation}\label{a13}
{k_i\over k_0}=-\sqrt{g\over f}~n^i=-{g\over f}~\!{k^i\over k^0}\;.
\end{equation}
From its definition (\ref{a11b}), and from the solution of the motion,
we find that the tangent vector ${\bf n}$ is given by

\begin{equation}\label{a14}
{\bf n} = \varepsilon_r\sqrt{1-{f\over g}{b^2\over r^2}}{\bf
e}_r+{b\over r}\sqrt{{f\over g}}{\bf e}_\alpha\;.
\end{equation} 
Next, the equations of motion for $r(t)$ and $\alpha (t)$ in the
orbital plane are

\begin{eqnarray}
{dr\over cdt}&=&\varepsilon_r {f\over g}\sqrt{{g\over f}-{b^2\over
r^2}}\label{a15a}\;,\\ r^2{d\alpha\over cdt}&=&b{f\over
g}\label{a15b}\;.
\end{eqnarray}
The differential equation for the trajectory is

\begin{equation}\label{a16}
{dr\over d\alpha}=\varepsilon_r r\sqrt{{g\over f}{r^2\over b^2}-1}\;.
\end{equation}

\noindent {\it Photon trajectory}.
The trajectory at the relativistic order $1/c^2$ (and even at order
$1/c^3$), which is the solution of (\ref{a16}), is an hyperbola whose
focus is the center of the Earth, with impact parameter $b$ and total
deviation angle $4GM/c^2 b$. The equation of the path reads

\begin{equation}\label{a17}
\cos (\alpha-\alpha_0) = {b\over r}-{2GM\over c^2b}\;,
\end{equation}
where the angle $\alpha_0$ represents the direction of the periastron,
at which the distance of closest approach is $r_0$, which, to order
$1/c^2$, is given by

\begin{equation}\label{a18}
r_0=b-{2GM\over c^2}\;.
\end{equation}
The tangent vector along the trajectory can be expressed in terms of
the vectors ${\bf e}_{\alpha_0}$ and ${\bf e}_{r_0}$, corresponding to
the position $(r_0,\alpha_0)$ of the periastron:

\begin{equation}\label{a19}
{\bf n}={\bf e}_{\alpha_0}-{2GM\over c^2 r_0}\sin (\alpha-\alpha_0)
{\bf e}_{r_0}\;.
\end{equation}
As for the radial equation (\ref{a15a}) it reads, after being
expressed in terms of $r_0$ rather than $b$,

\begin{equation}\label{a20}
c dt = \varepsilon_r {r dr\over \sqrt{r^2-r_0^2}} \left[1+{2GM\over
c^2r}\left(1+{r_0\over r+r_0}\right)\right]\;.
\end{equation}
This is readily integrated as

\begin{eqnarray}\label{a21}
c|t-t_0|&=&\sqrt{r^2-r_0^2}\nonumber\\&+&{2GM\over c^2}\Biggl[
\sqrt{{r-r_0\over r+r_0}}+\ln\biggl({r+\sqrt{r^2-r_0^2}\over
r_0}\biggr)\Biggr]\;,\nonumber\\
\end{eqnarray}
where $t_0$ denotes the instant of passage at the periastron. On the
other hand, the trajectory is obtained as 

\begin{equation}\label{a22}
|\alpha-\alpha_0|=\arctan\left({\sqrt{r^2-r_0^2}\over
r_0}\right)+{2GM\over c^2 r_0}\sqrt{{r-r_0\over r+r_0}}\;,
\end{equation}
where $\alpha_0$ is the angle at the periastron.  

\noindent {\it Coordinate time transfer}. We consider from now 
on a transfer from an emission point $A$ along the trajectory to some
reception point $B$. We introduce the {\it Euclidean} vectorial
distance between the two points, defined as the difference of the
coordinate positions of the points in the GRS coordinate system:

\begin{eqnarray}\label{a23}
{\bf R}_{AB}&=&{\bf r}_B-{\bf r}_A \;;~{\rm R}_{AB}=|{\bf R}_{AB}| \;;
~{\bf N}_{AB}={\bf R}_{AB}/{\rm R}_{AB}\;.\nonumber\\
\end{eqnarray}
The position of the emitting point $A$ is taken at the time of
emission $t_A$, ${\bf r}_A={\bf r}_A(t_A)$, and similarly ${\bf
r}_B={\bf r}_B(t_B)$. Notice that the norm ${\rm R}_{AB}$ is simply
the Euclidean norm; thus, ${\rm R}_{AB}$ represents the ``straight
line'' distance between $A$ and $B$. In practice it is very useful to
express all quantities in terms of such Euclidean notions of distance
and direction as ${\rm R}_{AB}$ and ${\bf N}_{AB}$. Once we have fixed
our coordinate system to be $(t,{\bf x})$, we can forget about the
curved geometry (\ref{a1}) and reason as if we were in flat
space-time.

Clearly, with the two points $A$ and $B$ being given, the trajectory
is entirely determined. In particular, the radial coordinate $r_0$ of
the periastron is uniquely fixed as a function of $r_A$, $r_B$ and
${\rm R}_{AB}$. Neglecting terms in $1/c^2$, we find

\begin{eqnarray}\label{a24}
r_0^2 = {1\over 4 {\rm R}_{AB}^2}\biggl({\rm
R}_{AB}^2-(r_A-r_B)^2\biggr)\biggl((r_A+r_B)^2-{\rm
R}_{AB}^2\biggr)\;.\nonumber\\
\end{eqnarray}
The unit vector ${\bf N}_{AB}$ is given as

\begin{equation}\label{a25}
{\bf N}_{AB}={\bf e}_{\alpha_0}-{2GM\over c^2 r_0}{r_B-r_A\over {\rm
R}_{AB}}{\bf e}_{r_0}\;.
\end{equation}
On the other hand, the unit tangent to the trajectory at the emission
point $A$ (say) is given from (\ref{a19}) as 

\begin{equation}\label{a26}
{\bf n}_A={\bf e}_{\alpha_0}-{2GM\over c^2 r_0}\sin
(\alpha_A-\alpha_0) {\bf e}_{r_0}\;.
\end{equation}
The difference between the two vectors (\ref{a25}) and (\ref{a26}) is
a small quantity ${\cal O}(1/c^2)$. It is not difficult, with the help
of our solution for the trajectory [{\it cf} (\ref{a17})-(\ref{a19})
and (\ref{a24})], to obtain the relation between these two vectors
(always working consistently to the order $1/c^2$):

\begin{eqnarray}\label{a27} 
{\bf n}_A&=&{\bf N}_{AB}\nonumber\\ &+&{4GM\over c^2 r_A}{{\rm
R}_{AB}\over (r_A+r_B)^2-{\rm R}_{AB}^2}\biggl({\bf r}_A-({\bf
N}_{AB}.{\bf r}_A){\bf N}_{AB}\biggr)\;.\nonumber\\
\end{eqnarray}
The coordinate time transfer from $A$ to $B$, denoted
$T_{AB}=t_B-t_A$, follows from the equation (\ref{a21}). By expressing
it with the help of the previous notation, notably of the ``straight
line'' distance ${\rm R}_{AB}$, we obtain the simple formula

\begin{equation}\label{a28}
T_{AB}={{\rm R}_{AB}\over c}+{2G M\over c^3} \ln\left(r_A+r_B+{\rm
R}_{AB}
\over r_A+r_B-{\rm R}_{AB}\right)\;. 
\end{equation}
The second term, purely of order $1/c^3$, is the Shapiro logarithmic
time delay.

The right side of (\ref{a28}) depends only on the spatial positions of
the emission and reception points, i.e. ${\bf r}_A$ and ${\bf r}_B$,
and not on the instants $t_A$ and/or $t_B$. Indeed, for stationary
space-times, the time $t-t_A$ elapsed from some emission instant $t_A$
does not depend on $t_A$ but only on the spatial coordinates ${\bf
r}_A$ and ${\bf r}$, so there exists a function $V$ such that
$c(t-t_A)=V({\bf r}_A,{\bf r})$. This function is the so-called
optical point characteristic of the stationary space-time, or optical
distance between pairs of points (Buchdahl 1970, 1979). In our case,
we have

\begin{equation}\label{a28'}
V({\bf r}_A,{\bf r})={\rm R}_{A}+{2G M\over c^2}
\ln\left(r_A+r+{\rm R}_{A}\over r_A+r-{\rm R}_{A}\right)\;,
\end{equation}
where ${\bf R}_A={\bf r}-{\bf r}_A$, ${\rm R}_A=|{\bf R}_A|$. Note
that along the light ray from ${\bf r}_A$ to ${\bf r}$, the function
$S$ defined by $S(t,{\bf r})= -ct+V({\bf r}_A,{\bf r})$ stays constant
($S=-ct_A$). Therefore $S$ represents the phase of the signal and can
be used to define the wave-vector as $k_\mu=\partial_\mu S$. For this
choice $k_0=-1$ and $k_i=\partial_iV$. Using (\ref{a28'}) we get

\begin{equation}\label{a28''}
{k_i\over k_0}=-{\rm N}_A^i-{4GM\over c^2}{(r_A+r){\rm N}_A^i-{\rm
R}_A{r^i\over r}\over (r_A+r)^2-{\rm R}_A^2}\;,
\end{equation}
where ${\rm N}_A^i={\rm R}_A^i/{\rm R}_A$. This is in complete
agreement with (\ref{a13}) and (\ref{a27}).

Given the simplicity of the result (\ref{a28}) for the time transfer
when expressed in terms of the Euclidean distance ${\rm R}_{AB}$, one
can guess that the formula can be derived directly by integrating
$ds^2=0$ along the path from $A$ to $B$. Let us show how this
works. With the post-Newtonian metric (\ref{a1})-(\ref{a3}) we have,
along the photon's path,

\begin{equation}\label{a29}
c~dt=\left(1+{2G M\over rc^2}\right)|d{\bf r}|\;,
\end{equation}
where $|d{\bf r}|$ is the Euclidean norm of the vector $d{\bf r}=
d{\bf x}$. Introducing ${\bf R}_A={\bf r}-{\bf r}_A$ as a new spatial
coordinate along the path, we have $|d{\bf r}|=(d{\rm R}_A^2+{\rm
R}_A^2d{\bf N}_A^2)^{1/2}$ where ${\rm R}_A=|{\bf R}_A|$ and ${\bf
N}_A={\bf R}_A/{\rm R}_A$. But we know from (\ref{a27}) that ${\bf
N}_A$ differs from the unit tangent ${\bf n}_A$ at the emission point
by a small term ${\cal O}(1/c^2)$. So, when ${\bf r}$ varies (the
origin point $A$ on the path being fixed), we have $d{\bf N}_A={\cal
O}(1/c^2)$ and therefore we see that $d{\bf N}_A^2={\cal O}(1/c^4)$
makes a negligible contribution. This shows that to order $1/c^2$
inclusively the time transfer can be calculated ``along the straight
line''; we have

\begin{equation}\label{a30}
dt=\left(1+{2G M\over rc^2}\right){d{\rm R}_A\over c}\;,
\end{equation}
and the total time transfer reads

\begin{equation}\label{a31}
T_{AB}={{\rm R}_{AB}\over c}+{2G M\over c^3} \int_0^{{\rm
R}_{AB}}{d{\rm R}_A\over r}\;.
\end{equation}
To find the closed-form expression of the integral, we insert $r=({\rm
R}_A^2+2~\!{\bf R}_A.{\bf r}_A+r_A^2)^{1/2}$ which can be
approximated, since ${\bf N}_A={\bf n}_A+{\cal O}(1/c^2)$ and the
integral already enters a small quantity, by $r=({\rm R}_A^2+2~\!{\rm
R}_A{\bf n}_A.{\bf r}_{A}+r_A^2)^{1/2}$. Next we perform the
integration over ${\rm R}_A$ from $A$ to $B$, and we are allowed to
replace ${\bf n}_A={\bf N}_{AB}+{\cal O}(1/c^2)$ within the
result. Finally we find (see e.g. Will 1981)

\begin{equation}\label{a32}
T_{AB}={{\rm R}_{AB}\over c}+{2G M\over c^3}
\ln\left(r_B+{\bf r}_B.{\bf N}_{AB}\over r_A+{\bf r}_A.
{\bf N}_{AB}\right)\;. 
\end{equation}
This expression for the Shapiro time delay is slightly different from
the previous form (\ref{a28}) but can easily be reconciled with it. To
this end one makes use of the identities

\begin{eqnarray}\label{a33}
r_B+{\bf r}_B.{\bf N}_{AB}&=&{(r_B+{\rm R}_{AB})^2-r_A^2\over 2{\rm
R}_{AB}}\;,\\ r_A+{\bf r}_A.{\bf N}_{AB}&=&{r_B^2-(r_A-{\rm
R}_{AB})^2\over 2{\rm R}_{AB}}\;,
\end{eqnarray}
which show that (\ref{a28}) and (\ref{a32}) are indeed totally
equivalent. However, in practice, we shall prefer to use the formula
(\ref{a28}) rather than (\ref{a32}) because of its structural
simplicity.

\noindent {\it Sagnac terms}. The formula (\ref{a28}) gives 
the time transfer from the point ${\bf r}_A\equiv {\bf r}_A(t_A)$ at
the emission instant $t_A$, to the reception point ${\bf r}_B\equiv
{\bf r}_B(t_B)$ at the reception instant $t_B$. But in fact, in the
application to ACES, the useful time basis in the experiment is that
provided by the clock at the point A (i.e., in the satellite) which
records the instant of emission $t_A$. Therefore, it is more
convenient to re-express the time transfer (\ref{a28}) in terms of the
position of the reception point $B$ as it was at the instant $t_A$
rather than at $t_B$, i.e. when it was instantaneous with the instant
of emission at $A$ (in the GRS coordinate system). Thus, we introduce
the ``instantaneous'' coordinate distance

\begin{equation}\label{a34}
{\bf D}_{AB}= {\bf r}_B(t_A)-{\bf r}_A(t_A)\ ;\quad {\rm D}_{AB}=|{\bf
D}_{AB}|\;,
\end{equation}
and perform a consistent series expansion when $1/c$ tends to zero. We
know from (\ref{a28}) that the time transfer $T_{AB}={\cal
O}(1/c)$. Therefore, with the required accuracy, we can write ${\bf
R}_{AB}={\bf D}_{AB}+T_{AB} {\bf v}_B(t_A)+{1\over 2} T_{AB}^2 {\bf
a}_B(t_A)+{\cal O}(1/c^3)$, where ${\bf a}_B$ is the acceleration. The
successive relativistic approximations are obtained by working out
this formula iteratively together with (\ref{a28}). In this way we
obtain

\begin{eqnarray}\label{a36}
T_{AB}&=&{{\rm D}_{AB}\over c}+{{\bf D}_{AB}.{\bf v}_B(t_A)\over
c^2}\nonumber\\ &+&{{\rm D}_{AB}\over 2c^3}\left({\bf v}_{B}^2+{({\bf
D}_{AB}.{\bf v}_B)^2\over {\rm D}_{AB}^2}+{\bf D}_{AB}.{\bf
a}_B\right)\nonumber\\ &+&{2GM\over c^3}\ln\left({r_A+r_B+{\rm
D}_{AB}\over r_A+r_B-{\rm D}_{AB}}\right)\;,
\end{eqnarray}
where all quantities are measured at the emission instant $t_A$
recorded by the on-board clock. The formula involves the Sagnac terms
of first ($1/c^2$) and second ($1/c^3$) orders, as well as the Shapiro
time delay [of course, consistently with the approximation, one can
use indifferently in the Shapiro term either the distance ${\rm
D}_{AB}$ or ${\rm R}_{AB}$, and either $r_B(t_B)$ or $r_B(t_A)$].

\noindent {\it One-way frequency transfer}. In the geometric 
optics approximation we have $k_\mu=\partial_\mu S$ and the phase
difference between two successive electromagnetic pulses is
$dS=(\partial_\mu S dx^\mu)_A=(\partial_\mu S dx^\mu)_B$, which in
other words means $(k_\mu u^\mu d\tau)_A=(k_\mu u^\mu d\tau)_B$ where
$u^\mu=dx^\mu/d\tau$ denotes the four-velocity (such that
$g_{\mu\nu}u^\mu u^\nu=-1$). Now, $d\tau_A$ and $d\tau_B$ are the
proper periods of the same signal at $A$ and $B$ and we have
$\nu_A/\nu_B=d\tau_B/d\tau_A$. Therefore, the one-way frequency shift
of photons during the transfer from $A$ to $B$ is

\begin{equation}\label{a37}
{\nu_A\over\nu_B}={(k_\mu u^\mu)_A\over (k_\mu u^\mu)_B}\;.
\end{equation}
Separating out temporal from spatial components, we have $k_\mu
u^\mu=u^0(k_0+k_i v^i/c)$, where $v^i=c u^i/u^0$ denotes the
coordinate velocity and where $u^0=(-g_{\rho\sigma}v^\rho
v^\sigma/c^2)^{-1/2}$. In a stationary space-time $k_0$ is constant
along the trajectory, i.e. $(k_0)_A=(k_0)_B=-{\hat E}/c$ in the
notation (\ref{a6a}); for instance, we have seen that with the choice
$S(t,{\bf r})= -ct+V({\bf r}_A,{\bf r})$, where $V$ is the optical
distance (\ref{a28'}), this constant is equal to minus one. Thus we
can write

\begin{equation}\label{a39}
{\nu_A\over\nu_B}={(u^0)_A\over (u^0)_B}~{1+\left({k_i\over
k_0}\right)_A{v_A^i\over c}\over 1+\left({k_i\over
k_0}\right)_B{v_B^i\over c}}\;.
\end{equation}
Next, we found in (\ref{a13}) that the covariant components of the
wave vector $k_i$ are proportional to the unit tangent vector
$n^i$. Using this we can further infer that

\begin{equation}\label{a40}
{\nu_A\over\nu_B}={(u^0)_A\over (u^0)_B}~{1-\sqrt{{g(r_A)\over
f(r_A)}}{{\bf n}_A.{\bf v}_A\over c} \over 1-\sqrt{{g(r_B)\over
f(r_B)}}{{\bf n}_B.{\bf v}_B\over c}}\;.
\end{equation}
Within the first factor $u^0$ is the coordinate time {\it vs} proper
time ratio, which has been computed in equation (\ref{3}). Thus,
from $(u^0)_A/(u^0)_B=(d\tau/dt)_B/(d\tau/dt)_A$, we get

\begin{equation}\label{a41}
{(u^0)_A\over (u^0)_B}={1-{1\over c^2}\left[U_{\rm E}({\bf r}_B)+{{\bf
v}_B^2\over 2}\right]\over 1-{1\over c^2}\left[U_{\rm E}({\bf
r}_A)+{{\bf v}_A^2\over 2}\right]}\;,
\end{equation}
where $U_A$ and $U_B$ are the Newtonian potentials at the levels $A$
and $B$. The factor (\ref{a41}) comprises the Einstein gravitational
red-shift and the second-order Doppler effects, which are both of
order $1/c^2$. We recall from Section \ref{sect2} that at the level of
accuracy of ACES, the potentials in (\ref{a41}) must take into account
the oblaticity of the Earth.

Now the second factor in the right-hand-side of (\ref{a40}) is nothing
but the ratio of coordinate periods of the same signal at $A$ and $B$,
namely $dt_B/dt_A$; it contains the first-order Doppler effect ($\sim
1/c$) and the third-order ($\sim 1/c^3$) terms we are looking for. For
the computation of this factor at the level $5\times 10^{-17}$ we do not
need to consider the $J_2$ of the Earth potential, and we can
approximate $U_{\rm E}=GM/r$. In order to obtain a useful formula, we
have only to substitute for the tangent vectors ${\bf n}_A$ and ${\bf
n}_B$ their expressions in terms of the unit direction ${\bf N}_{AB}$
into (\ref{a40}). The required relation between these vectors was
found in (\ref{a27}). Alternatively, one can use directly the formula
(\ref{a28''}). Finally, our end-result for the one-way frequency
transfer with $1/c^3$ accuracy reads

\begin{equation}\label{a42}
{\nu_A\over\nu_B}={(u^0)_A\over (u^0)_B}~{q_A
\over q_B}\;, 
\end{equation}
where the ratio $q_A/q_B$ is given by the formulas
(\ref{13a})-(\ref{13b}) in the text.

The previous expressions were obtained using our explicit solution
(\ref{a17})-(\ref{a19}) for the photon's trajectory. However, there is
a simpler way to obtain the ratio $q_A/q_B$, which is based on the
fact that since it is equal to the ratio of coordinate times:
$q_A/q_B=dt_B/dt_A$, it can be computed directly by differentiating
the coordinate time transfer $T_{AB}=t_B-t_A$ with respect to the
emission time $t_A$. The time transfer to order $1/c^3$ is given by
the simple formula (\ref{a28}) containing the Shapiro term. Thus, we
must consider

\begin{eqnarray}\label{a44}
&&{d\over dt_A}\left(t_B-t_A\right)=\nonumber\\ &&\quad {d\over
dt_A}\left\{{{\rm R}_{AB}\over c}+{2G M\over c^3}
\ln\left(r_A+r_B+{\rm R}_{AB}\over r_A+r_B-{\rm R}_{AB}\right)\right\}\;.
\end{eqnarray}
The differentiation is to be performed taking into account the fact
that the coordinate distance between $A$ and $B$ depends on both the
emission and reception times, i.e. ${\rm R}_{AB}=|{\bf r}_B(t_B)-{\bf
r}_A(t_A)|$. Thus we have for instance

\begin{equation}\label{a45}
{d{\rm R}_{AB}\over dt_A}={\bf N}_{AB}.\left({\bf v}_B{dt_B\over
dt_A}-{\bf v}_A\right)\;.
\end{equation}
We find that the ratio $q_A/q_B=dt_B/dt_A$ as obtained from the
equation (\ref{a44}) agrees exactly with the one given by the
expressions (\ref{13a})-(\ref{13b}).

\noindent {\it Two-way frequency transfer}. In this method 
the ``tracking'' signal is emitted from the ground station at point
$B'$ and coordinate time $t_{B'}$, is received by the satellite at
point $A$ and time $t_A$ and immediately transponded back to the
ground station where it arrives at point $B$ and time $t_B$. (A
generalization of the formulae to the case where there is a time delay
between reception and re-emission at $A$ is also possible.) The
``clock'' signal is emitted at $A$, simultaneously with the tracking
signal when it is transponded at $A$, and transmitted to $B$. This
technique, called the Doppler-cancelling technique, permits us to
suppress the first-order Doppler effect and to drastically limit the
uncertainties linked with the atmospheric contributions (see
e.g. Kleppner, Vessot \& Ramsey 1970). The ratio of the signal
frequencies to be inserted in (\ref{11}) follows from

\begin{equation}\label{a46}
{\nu_B\over \nu_A}={1\over 2}
\left({\nu_B\over \nu_{B'}}\right)_{\rm station}+\Delta_{AB}+{1\over 2}\;,
\end{equation}
where $(\nu_B/\nu_{B'})_{\rm station}$ is measured at the ground
station, and where $\Delta_{AB}$ is given theoretically by

\begin{equation}\label{a47}
\Delta_{AB}=\left({\nu_B\over \nu_A}\right)\left[1-{1\over 2}
\left({\nu_A\over \nu_{B'}}\right)\right]-{1\over 2}\;.
\end{equation}
The ratios $\nu_B/\nu_A$ and $\nu_A/\nu_B'$ are given by the formulas
(\ref{12})-(\ref{13b}), valid for the one-way transfer, that we must
simply insert into (\ref{a47}) in order to get the result. The only
problem is to correctly express all the quantities associated with the
up-link from the ground station $B'$ at $t_{B'}$ to the satellite $A$
at $t_A$, in terms of the same quantities associated with the down-link
back from the satellite $A$ to the station $B$ at $t_B$ (see Figure
\ref{fig2}). Notably we need to express the unit vector ${\bf
N}_{B'A}$ in terms of the unit vector ${\bf N}_{AB}$ on the way back,
as well as the velocity and acceleration ${\bf v}_B(t_B)$ and ${\bf
a}_B(t_B)$ of the ground station at $t_B$. Neglecting ${\cal
O}(1/c^3)$ we find

\begin{eqnarray}\label{a48}
{\bf N}_{B'A}=&-&{\bf N}_{AB}\biggl\{1+{2\over c}{\bf N}_{AB}.{\bf
v}_B\nonumber\\ &+&{1\over c^2}\left[4({\bf N}_{AB}.{\bf v}_B)^2-2{\bf
v}_B^2-2{\bf R}_{AB}.{\bf a}_B\right]\biggr\}\nonumber\\ &+&{2\over
c}{\bf v}_B\left(1+{1\over c}{\bf N}_{AB}.{\bf v}_B\right)-{2\over
c^2}{\rm R}_{AB} {\bf a}_B\;.
\end{eqnarray}
We need also the velocity of the station at emission, ${\bf
v}_{B'}(t_{B'})$, in terms of its velocity, acceleration and
derivative of acceleration at reception, ${\bf v}_{B}(t_{B})$, ${\bf
a}_{B}(t_{B})$ and ${\bf b}_{B}(t_{B})$. We get

\begin{eqnarray}\label{a49}
{\bf v}_{B'}&=&{\bf v}_B-{2\over c}{\rm R}_{AB}{\bf a}_B+{2\over
c^2}\left[({\bf R}_{AB}.{\bf v}_B){\bf a}_B+{\rm R}_{AB}^2{\bf
b}_B\right]\;.\nonumber\\
\end{eqnarray}
These relativistic expansions being fully taken into account, we
obtain finally, after a long computation, the formula (\ref{15}) in
the text, which is

\begin{eqnarray}\label{a50}
\Delta_{AB}\!&=&\!{1\over c^2}\!\left[U_{AB}-{1\over 2}
{\bf v}_{AB}^2-{\bf R}_{AB}.{\bf a}_B\right]\!\left(1+{{\bf
N}_{AB}.{\bf v}_{AB}
\over c}\right)\nonumber\\
&+&{{\rm R}_{AB}\over c^3}\left(-{\bf v}_A.{\bf a}_B+{\bf R}_{AB}.{\bf
b}_B+2{\bf v}_B.{\bf a}_B-{\bf v}_B.\nabla U_B\right)\;,\nonumber\\
\end{eqnarray}
where $U_{AB}=U_B-U_A$ and ${\bf v}_{AB}={\bf v}_A-{\bf v}_B$. All the
quantities at $A$ or $B$ are expressed at the corresponding instants
$t_A$ or $t_B$ respectively.

\end{document}